# Rotary dipole-mode solitons in Bessel photonic lattices

Yaroslav V. Kartashov,[1,2] Alexey A. Egorov,[2] Victor A. Vysloukh,[3] Lluis Torner[1]

[1]*ICFO-Institut de Ciencies Fotoniques and Department of Signal Theory and Communications, Universitat Politecnica de Catalunya, 08034 Barcelona, Spain*
[2]*Physics Department, M. V. Lomonosov Moscow State University, 119899 Moscow, Russia.*
[3]*Departamento de Fisica y Matematicas, Universidad de las Americas, 72820 Puebla, Mexico.*

We address Bessel photonic lattices of radial symmetry imprinted in cubic Kerr-type nonlinear media and show that they support families of stable dipole-mode solitons featuring two out-of-phase light spots located in different lattice rings. We show that the radial symmetry of the Bessel lattices afford a variety of unique soliton dynamics including controlled radiation-free rotation of the dipole-mode solitons.

*PACS numbers: 42.65.Wi, 42.65.Jx*

*Keywords: optical lattices, solitons*

The basic features of propagation of optical radiation in media with transverse modulation of the refractive index are known to depart considerably from those for uniform media. Local refractive index maxima with appropriate characteristics forms waveguides that enable self-trapping and guiding of light beams. When the refractive index is modulated periodically, thus creating array of evanescently coupled waveguides, the formation of discrete solitons is possible in the neighboring array sites [1,2]. Discrete solitons exhibit a number of unique features including controllable steering and switching that makes them promising to demonstrate all-optical routing concepts [3]. Here we study the intermediate regime encountered in lattices with variable refractive index modulation depth, capable of operating in both regimes of weak and strong coupling between neighboring sites [4,5]. The concept behind such regime might be termed *tunable discreteness*, with the strength of modulation being the parameter that tunes the system properties from predominantly continuous to predominantly discrete. In recent landmark experiments it was showed that lattices with flexibly controlled refractive index modulation period and depth can be induced optically [6-10], which exhibits a number of advantages in comparison with technological fabrication of the evanescently coupled waveguide arrays. The properties of lattice solitons supported by such kind of periodic lattices are being comprehensively investigated (see Refs [4-15]).

The possibilities afforded by the lattices for soliton control depend crucially on the lattice overall symmetry. In particular, the periodicity of lattices dictates the band-gap structure of the existence domain for lowest-order solitons and their low-power asymptotical profiles (Bloch waves); it also defines the permitted configurations of soliton arrays, and the symmetry of the corresponding field distributions (e.g., quasi-discrete vortices). Therefore, a new wealth of possibilities is open, in both optics and matter wave fields, by lattices of radial symmetry created with non-diffracting Bessel beams. Those feature a cylindrical symmetry, thus new phenomena non-accessible with periodic (pixel-like) lattices. In this paper we address unique possibilities afforded by the Bessel lattices. In particular we introduce families of dipole-mode spatial solitons supported by Bessel photonic lattice imprinted in focusing cubic nonlinear medium and perform their stability analysis. We show that such dipole-mode solitons are stable under propagation above a minimum power threshold, and we illustrate the concept of the controlled rotation and reorientation of the dipole-mode solitons by an auxiliary control beam.

We consider propagation of optical radiation along $z$ axis in a bulk cubic medium with transverse modulation of linear refractive index described by nonlinear Schrödinger equation for dimensionless complex field amplitude $q$:

$$i\frac{\partial q}{\partial \xi} = -\frac{1}{2}\left(\frac{\partial^2 q}{\partial \eta^2} + \frac{\partial^2 q}{\partial \zeta^2}\right) + \sigma q |q|^2 - pR(\eta,\zeta)q. \qquad (1)$$

Here longitudinal $\xi$ and transverse $\eta,\zeta$ coordinates are scaled to the diffraction length and input beam width, respectively, $\sigma = \mp 1$ stands for focusing/defocusing nonlinearity. Parameter $p$ is proportional to the depth of refractive index modulation. Here we consider the case of simplest radial Bessel lattice described by $R(\eta,\zeta) = J_0[(2b_{\text{lin}})^{1/2}(\eta^2 + \zeta^2)^{1/2}]$, where $b_{\text{lin}}$ defines radiuses of lattice rings, and leave more complicated lattices, including those with azimuthal modulation, for future consideration. Notice that the function $q(\eta,\zeta,\xi) = J_0[(2b_{\text{lin}})^{1/2}(\eta^2 + \zeta^2)^{1/2}]\exp(-ib_{\text{lin}}\xi)$ gives exact solution of linear homogeneous Eq. (1) at $\sigma = p = 0$, and describes non-diffracting two-dimensional laser beams, in close analogy with intersecting planar waves used for lattice formation in Refs [6-10]. We assume that the depth of refractive index modulation is small compared with the unperturbed refractive index and is of the order of the nonlinear contribution. Notice that Eq. (1) admits several conserved quantities including the energy flow $U$, the Hamiltonian $H$, and the longitudinal projection of the angular momentum $\mathbf{L}$:

$$U = \int_{-\infty}^{\infty}\int_{-\infty}^{\infty} |q|^2 \, d\eta d\zeta,$$
$$H = \int_{-\infty}^{\infty}\int_{-\infty}^{\infty} \left(\frac{1}{2}|\nabla q|^2 + \frac{\sigma}{2}|q|^4 - pR|q|^2\right) d\eta d\zeta, \qquad (2)$$
$$L_\xi = \mathbf{L}\mathbf{e}_\xi = \frac{\mathbf{e}_\xi}{2i}\int_{-\infty}^{\infty}\int_{-\infty}^{\infty} [\mathbf{r} \times (q^*\nabla q - q\nabla q^*)]d\eta d\zeta.$$

Here $\nabla = \mathbf{e}_\eta(\partial/\partial\eta) + \mathbf{e}_\zeta(\partial/\partial\zeta)$ is the differential operator acting in the transverse plane, $\mathbf{r} = \mathbf{e}_\eta\eta + \mathbf{e}_\zeta\zeta$ is the radius vector of the point with coordinates $(\eta,\zeta)$, $\mathbf{e}_\eta, \mathbf{e}_\zeta, \mathbf{e}_\xi$ are unitary vectors in the directions $\eta,\zeta,\xi$, respectively.

We search for soliton solutions in the form $q(\eta,\zeta,\xi) = w(\eta,\zeta)\exp(ib\xi)$, where $w(\eta,\zeta)$ is a real function and $b$ is a real propagation constant. Soliton families are defined by the propagation constant $b$, the parameter $b_{\text{lin}}$ and the lattice depth $p$. Since scaling transformations $q(\eta,\zeta,\xi,p) \to \chi q(\chi\eta,\chi\zeta,\chi^2\xi,\chi^2 p)$ can be used to obtain various families of lattice solitons from a given one, here we set the transverse scale in such way that $b_{\text{lin}} = 10$, and vary $b$ and $p$. In this paper we concentrate only on focusing nonlinear media with $\sigma = -1$, although we verified that defocusing Bessel lattices also support localized solitons. Linear stability analysis of the solitons was carried out by searching for the perturbed solutions $q(\eta,\zeta,\xi) = [w(\eta,\zeta) + u(\eta,\zeta,\xi) + iv(\eta,\zeta,\xi)]\exp(ib\xi)$, with $u$ and $v$ being the real and imaginary parts of the perturbation which can grow with complex growth rate $\delta$. A standard linearization procedure for Eq. (1) yields a system of coupled equations for the perturbation components $u, v$:

$$\frac{\partial u}{\partial \xi} = -\frac{1}{2}\left(\frac{\partial^2 v}{\partial \eta^2} + \frac{\partial^2 v}{\partial \zeta^2}\right) + bv - w^2 v - pRv,$$
$$-\frac{\partial v}{\partial \xi} = -\frac{1}{2}\left(\frac{\partial^2 u}{\partial \eta^2} + \frac{\partial^2 u}{\partial \zeta^2}\right) + bu - 3w^2 u - pRu. \quad (3)$$

This system was solved numerically with a split-step Fourier method in order to find the perturbation profiles and the corresponding growth rates.

First we consider dipole-mode solitons that intuitively can be viewed as nonlinear superposition of out-of-phase soliton beams supported by the central core and the first lattice ring. Because of the repulsive interaction between the out-of-phase beams, such localized states can not exist in the absence of Bessel lattice. The lattice compensates the repulsion and leads to stationary propagation. The typical profile of a stationary dipole-mode soliton calculated from Eq. (1) by relaxation method is presented in Fig. 1(a). From now on, these solutions will be referred to as centered dipole-mode solitons. The energy flow of centered dipole-mode soliton is nonmonotonic function of the propagation constant $b$, as shown in Fig. 1(b). At high energy flows dipole-mode soliton transforms into a pair of narrow non-interacting bright solitons. When $b \to \infty$, the total energy flows tends to twice the energy of the so-called Townes soliton. There exist a lower cutoff $b_{co}$ on propagation constant that is a non-monotonic function of lattice depth (Fig. 1(c)). Close to the lower cutoff the dipole-mode solitons become spatially extended and cover several lattice rings. At approximately $p \leq 8.2$ centered dipole-mode solitons cease to exist at lower cutoff without any topological transformation, while at $p > 8.2$ it transforms into radially-symmetric solutions that appears as a bright central peak surrounded by concentric rings with the shape of the linear guided mode of radially symmetric Bessel lattice. This transformation of the asymptotical soliton shapes corresponds to the discontinuity in the $b_{co}(p)$ dependence (Fig. 1(c)). Notice that with growth of $p$ the part of soliton energy carried out by the core decreases and at $p \to \infty$ most of the soliton energy is concentrated in the first lattice ring.

Our stability analysis revealed that centered dipole-mode solution becomes free of linear instabilities above certain energy flow threshold (above the corresponding critical value of propagation constant). The instability domain is located near the cutoff and the corresponding perturbation growth is of oscillatory type (Fig. 1(d)). Our calculations reveal that the width of the instability region reduces when the lattice depth grows. Close to cutoff we found complex narrow stability bands, whose detailed study falls beyond the scope of this paper, thus we do not display them in Fig. 1(d).

Besides the centered dipole-solitons discussed above, we have found a new type of dipole-mode solitons, which are supported by the first lattice ring (Fig. 2(a)). Those will be referred to as ring dipole-mode solitons. At approximately $p > 7.5$ such solitons transform into linear waves at lower cutoff and their energy flow vanishes, while at $p \leq 7.5$ ring dipole-mode ceases to exist at cutoff without any topological transformation (Fig. 2(b)). As in the case of centered solitons this is accompanied by the discontinuity of the $b_{co}(p)$ dependence (Fig. 2(c)). At high energy flows the two beams forming the ring dipole-mode solitons become very narrow, while at low energies they spread over the first lattice ring (Fig. 1(a)), so that their overlap becomes high. The same behavior is encountered with growth of the lattice depth at fixed energy flow.

The linear stability analysis shows that ring dipole-mode solitons also turn to be completely stable above energy flow threshold (Fig. 2(d)). Instability band is located near lower cutoff and its width decreases with growth of the lattice depth so that at $p = 20$ ring dipole-mode solitons become stable almost in the entire domain of their existence. It should be mentioned that at critical for stabilization value of propagation constant, the two beams forming the ring dipole-mode soliton still overlap considerably. This indicates that stabilization is due to the lattice. Notice that, inside the instability bands, both oscillatory and exponential instabilities occur with ring dipole-mode solitons.

The central new possibilities open by the Bessel lattices are linked to their radial symmetry, which makes possible to set solitons, and in particular dipole-mode solitons, into controlled rotation. Rotation of the so-called centered dipole-mode solitons can be achieved by launching one component trapped in first lattice ring tangentially to the ring at an angle $\alpha$. Ring dipole-mode solitons can be set into rotary motion by imposing on them an initial phase twist $\exp(i\nu\phi)$, where $\phi$ is the azimuth angle and $\nu$ is stands for the so-called winding number or phase twist (Fig. 3). In contrast to solitons supported by periodic lattices, that always radiate energy upon its motion across the lattice because they ought to overcome opposing potential across the lattice,[3] rotating dipole-mode solitons in Bessel lattices do not radiate. This property of solitons supported by Bessel lattices gives rise for a number of unique interaction types.

For example, the symmetry-axis of the dipole-mode solitons can jump, e.g., upon the action of an additional control beam launched into the second lattice ring. If the energy flow of the control beam largely exceeds that of dipole-mode solitons, the control beam does not experience considerable displacement upon interaction, while the axis of dipole-mode solitons can rotate for angles up to 180 degrees. This concept is illustrated in Fig. 4, where the arrows show the direction of reorientation of the dipole-mode soliton axis and the direction of displacement of the control beam. The control beam is out-of-phase with the dipole-mode soliton located in the first lattice ring. The direction of the rotation and its rate are dictated by the position of the control beam, and by the relative phase between the control beam and the dipole-mode soliton. . Notice that nonlinearity is necessary for the observed behavior throughout the paper, because even though linear guided modes analogous to the ring dipole-mode solitons that we describe here can in principle be supported by linear Bessel lattices, their rotation is always accompanied by radiation losses.

In conclusion, we have revealed properties of two types of dipole-mode solitons supported by radially symmetric Bessel photonic lattices in Kerr-type cubic nonlinear media. We showed that such solitons become completely stable above a minimum power threshold. Because of the radial symmetry of the lattice, we showed that the solitons can be set into controlled, radiation-free rotation, thus featuring a number of unique soliton dynamics.

This work has been partially supported by the Generalitat de Catalunya, and by the Spanish Government through grant BFM2002-2861.

**Figure captions**

Figure 1.  (a) Profile of centered dipole-mode soliton corresponding to the point marked by circle in dispersion diagram (b). (c) Cutoff on propagation constant versus lattice depth. (d) Real part of perturbation growth rate versus propagation constant at $p=9$.

Figure 2.  (a) Profile of ring dipole-mode soliton corresponding to point marked by circle in dispersion diagram (b). (c) Cutoff on propagation constant versus lattice depth. (d) Real part of perturbation growth rate versus propagation constant at $p=6$.

Figure 3.  Induced rotation of centered (a) and ring (b) dipole-mode solitons in Bessel lattice. Soliton shown in row (a) corresponds to $b=4$ at $p=9$. Soliton component trapped in first lattice ring is set in rotary motion by launching it tangentially to lattice ring at an angle $\alpha=0.1$. Soliton shown in row (b) corresponds to $b=3$ at $p=20$. Both components are set in rotary motion by imposing initial phase twist $\nu=0.1$.

Figure 4.  Reorientation of centered (a) and ring (b) dipole-mode solitons under the action of high-energy control soliton launched into second lattice ring. Centered and ring dipole-mode solitons correspond to $b=3$. High-energy control soliton is taken at $b=10$. Lattice depth $p=9$.

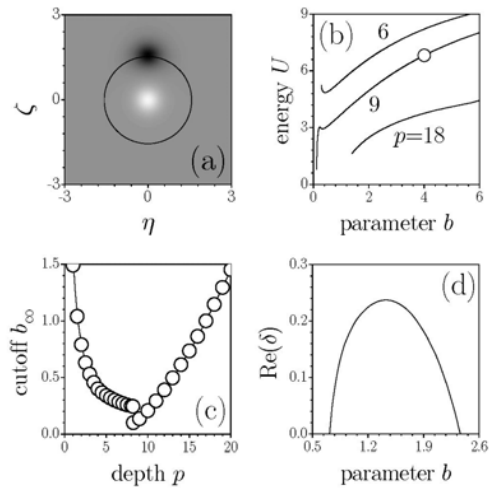

Figure 1. (a) Profile of centered dipole-mode soliton corresponding to the point marked by circle in dispersion diagram (b). (c) Cutoff on propagation constant versus lattice depth. (d) Real part of perturbation growth rate versus propagation constant at $p=9$.

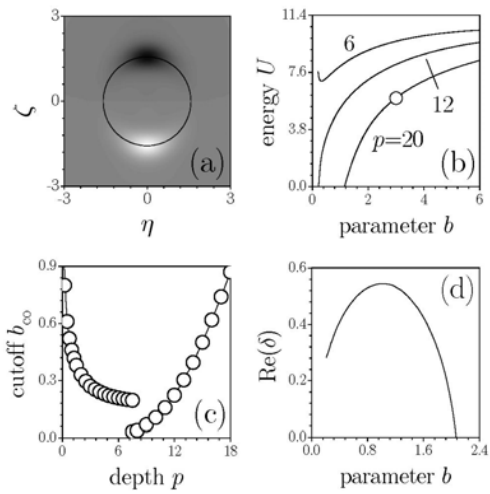

Figure 2. (a) Profile of ring dipole-mode soliton corresponding to point marked by circle in dispersion diagram (b). (c) Cutoff on propagation constant versus lattice depth. (d) Real part of perturbation growth rate versus propagation constant at $p=6$.

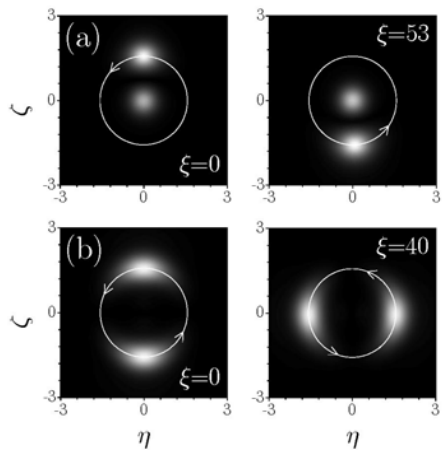

Figure 3.  Induced rotation of centered (a) and ring (b) dipole-mode solitons in Bessel lattice. Soliton shown in row (a) corresponds to $b=4$ at $p=9$. Soliton component trapped in first lattice ring is set in rotary motion by launching it tangentially to lattice ring at an angle $\alpha=0.1$. Soliton shown in row (b) corresponds to $b=3$ at $p=20$. Both components are set in rotary motion by imposing initial phase twist $\nu=0.1$.

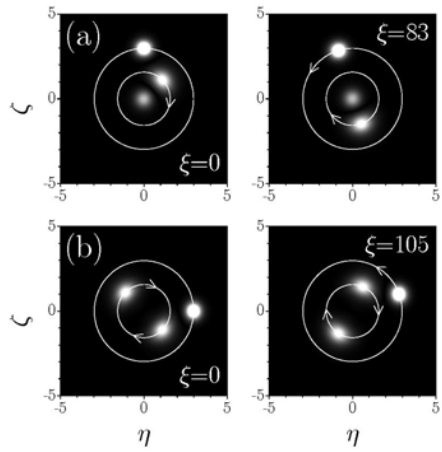

Figure 4. Reorientation of centered (a) and ring (b) dipole-mode solitons under the action of high-energy control soliton launched into second lattice ring. Centered and ring dipole-mode solitons correspond to $b = 3$. High-energy control soliton is taken at $b = 10$. Lattice depth $p = 9$.